\documentclass{aa}
\usepackage{booktabs}
\usepackage{graphicx}
\usepackage{longtable}
\usepackage{supertabular}

\def\simgt{\ {\raise-.5ex\hbox{$\buildrel>\over\sim$}}\ }
\def\simlt{\ {\raise-.5ex\hbox{$\buildrel<\over\sim$}}\ }

\def\cd{d$^{-1}$\,}
\def\kms{km\,s$^{-1}$\,}
\def\hlinewd#1{\noalign{\ifnum0=`}\fi\hrule \@height #1 \futurelet\reserved@a\@xhline} 
\begin{document}

\title{Time-resolved multicolour photometry of bright B-type variable
stars in Scorpius\,\thanks{Based on observations carried out at the South
African Astronomical Observatory}}
\authorrunning{Handler \& Schwarzenberg-Czerny}
\titlerunning{Bright B-type variables in Scorpius}
   \author{G. Handler, A. Schwarzenberg-Czerny}
   \offprints{G. Handler}

   \institute{
Copernicus Astronomical Center, Bartycka 18, 00-716 Warsaw, Poland (gerald@camk.edu.pl)}

\date{Received May 13, 2013; Accepted July 8, 2013}

\abstract
{The first two of a total of six nano-satellites that will constitute the
BRITE-Constellation space photometry
mission have recently been launched successfully.} 
{In preparation for this project, we carried out time-resolved colour 
photometry in a field that is an excellent candidate for BRITE
measurements from space.}
{We acquired 117 h of Str\"omgren $uvy$ data during 19 nights. Our targets 
comprised the $\beta$ Cephei stars $\kappa$ and $\lambda$ Sco, the 
eclipsing binary $\mu^1$ Sco, and the variable super/hypergiant 
$\zeta^1$ Sco.}
{For $\kappa$ Sco, a photometric mode identification in combination with 
results from the spectroscopic literature suggests a dominant $(l, m) = 
(1, -1)$ $\beta$~Cephei-type pulsation mode of the primary star. The longer 
period of the star may be a rotational variation or a g-mode pulsation.  
For $\lambda$ Sco, we recover the known dominant $\beta$ Cephei 
pulsation, a longer-period variation, and observed part of an eclipse. 
Lack of ultraviolet data precludes mode identification for this star. We 
noticed that the spectroscopic orbital ephemeris of the closer pair in 
this triple system is inconsistent with eclipse timings and propose a 
refined value for the orbital period of the closer pair of 
$5.95189\pm0.00003$\,d. We 
also argue that the components of the $\lambda$ Sco system are some 30\% 
more massive than previously thought. The binary light curve solution of 
$\mu^1$ Sco requires inclusion of the irradiation effect to explain the 
$u$ light curve, and the system could show additional low amplitude 
variations on top of the orbital light changes. $\zeta^1$ Sco shows 
long-term variability on a time scale of at least two weeks that we 
prefer to interpret in terms of a variable wind or strange mode 
pulsations.}
{}


\maketitle

\section{Introduction}

BRIght Target Explorer (BRITE)-Constellation is a space photometry 
mission consisting of six nano-satellites, two of each funded by the 
partner countries Austria, Canada, and Poland. The two Austrian 
satellites have been successfully launched on 25 February 2013, and the 
remaining four should sequentially follow until early 2014. The goal of 
the mission is to obtain high-quality time-resolved photometry of the 
brightest stars in the sky, in carefully selected fields with a diameter 
of 24\degr. Subrasters of the CCD images of up to a dozen primary 
targets will be transmitted to Earth, and on-board reduced photometry 
can be obtained for up to a hundred further targets in the field. All 
satellites carry a 3-cm telescopes, but three of them will observe 
through a blue filter, and the other three through a red filter 
(approximating the Johnson B and R passbands). This setup will allow to 
obtain two-colour photometry, and the satellites can be operated either 
separately, or as an entity, depending on the science needs. We refer to 
Zwintz \& Kaiser (2008) for more details.

Given the technical setup of the mission, it is expected to deliver the 
richest science return for OB stars. Being intrinsically luminous, such 
stars also feature prominently among the apparently brightest stars. OB 
stars naturally occur in groups, associations and clusters. Therefore 
sky fields rich in bright stars will also be rich in OB stars.

Variability is common among such stars, starting from the slowly 
pulsating B (SPB) and $\beta$
Cephei pulsators on the main sequence, extending to pulsating supergiants,
Luminous Blue Variables, Be stars, massive eclipsing binaries, etc. The
prime observable that supplies astrophysical information of a variable
star is the period of variability. The presence of multiperiodicity and
the shape of the light curves are also revealing. However, time-resolved
colour photometry provides valuable additional astrophysical information,
namely on the presence and nature of temperature variations, that is
essential for interpreting the observed variability. A prominent example
is the ability to identify the spherical harmonics of pulsation modes
(e.g., see Daszy{\'n}ska-Daszkiewicz 2008 in the context of
BRITE-Constellation).


In preparation for BRITE-Constellation, we carried out time-resolved 
multicolour photometry of bright early B-type stars in a field deemed to 
be likely observed by the mission. We chose a field in Scorpius, that 
contains the two $\beta$~Cephei stars $\kappa$ Sco ($V=2.4$) and 
$\lambda$ Sco ($V=1.6$), the eclipsing binary $\mu^1$ Sco ($V=3.1$), the 
variable $\zeta^1$~Sco ($V=4.7$), and the two apparently constant stars 
$\upsilon$ Sco ($V=2.7$) and $\mu^2$ Sco ($V=3.6$) for comparison 
purposes. All these stars are located within 10\degr in the sky and 
their magnitude range is what can be expected within a typical 
BRITE-Constellation field.

\section{Observations}

\begin{figure*}[htb]
\includegraphics[angle=270,width=180mm]{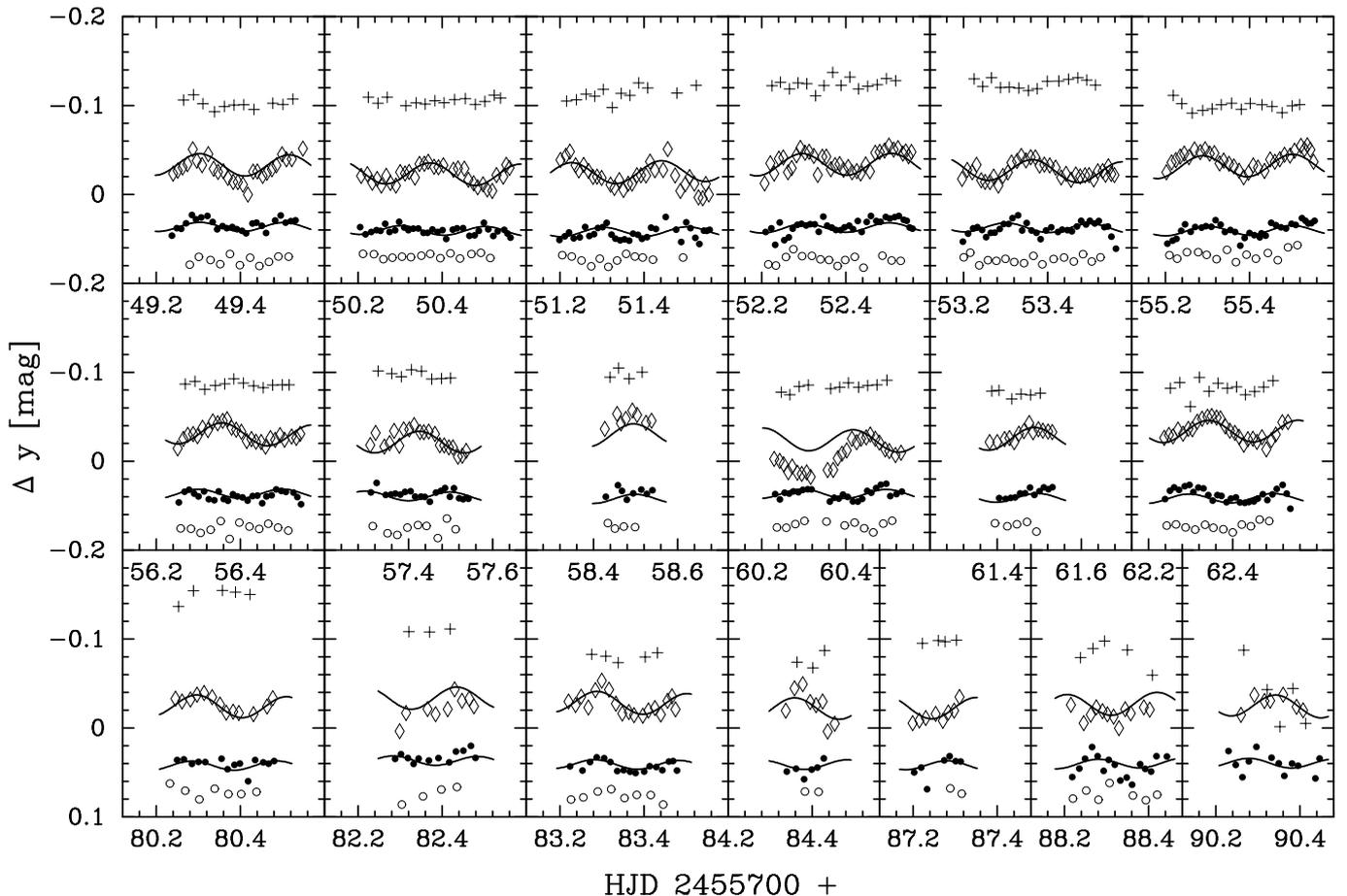}
\caption{Collected y-filter light curves of the low-amplitude and 
non-variables. Plus signs: $\zeta^1$~Sco. Diamonds: $\lambda$~Sco.
Filled circles: $\kappa$~Sco. Open circles: $\mu^2$ Sco $-\upsilon$ Sco.
The fits plotted over the $\kappa$~Sco and $\lambda$~Sco light curves 
are derived in Sect. 3.1 and 3.2.}
\end{figure*}

Time-resolved photometry was carried out at the 0.5-m telescope at the
South African Astronomical Observatory. We were awarded two runs of two
weeks each, separated by two weeks, in July and August 2011. The
photoelectric Modular Photometer that is permanently mounted to this
telescope was used in connection with Str\"omgren $uvy$ and neutral
density filters. For the five ``fainter" stars, the 2.3 mag neutral
density filter was sufficient, but for $\lambda$~Sco, the 4.7 mag neutral
density filter had to be used. Unfortunately, this filter has zero
transmission in the ultraviolet, wherefore we could only obtain $vy$
photometry for this star.

Because the two known $\beta$~Cephei stars exhibit faster light variations
than the other variables, the group $\kappa$/$\lambda$/$\upsilon$ Sco was
observed twice as often as the $\mu^1$/$\mu^2$/$\zeta^1$ Sco ensemble. Sky
background was measured at least once per group, depending on the position
and brightness of the Moon. This resulted in an average cadence of 17
minutes per measurement for the first group, and about 33 minutes per
measurement for the second. In total, we acquired 117 hours of measurement
during 19 nights.

The data were reduced following standard photoelectric photometry 
schemes. First, the measurements were corrected for coincidence losses. 
Then, sky background was subtracted within each target/local comparison 
star group, and if found necessary, individually for each star. 
Extinction coefficients were determined on a nightly basis from the 
measurements of the two comparison stars via the Bouguer method (fitting 
a straight line to a magnitude vs.\ air mass plot). The same extinction 
correction was applied to each star. Differential magnitudes relative to 
the two comparison stars were computed by interpolation. The magnitude 
differences between the comparison stars themselves had standard 
deviations of 6.8, 6.5, and 5.3 mmag in $u$, $v$ and $y$, respectively, 
and no evidence for periodic variability of any of these two stars was 
found within the (rather unsatisfactory) limitations of our data. 
Finally, the timings were converted to Heliocentric Julian Date (HJD), 
and the data for the variable stars were subjected to analysis. 
Differential light curves in the y filter are displayed in Fig.\ 1, 
excluding $\mu^1$ Sco whose amplitude is much larger than that of the 
other variables.

\section{Analysis, results and discussion}

The data were searched for periodicities using the program {\tt 
Period04} (Lenz \& Breger 2005). This software package uses single 
frequency Fourier and multifrequency nonlinear least squares fitting 
algorithms. The analysis was started by computing the spectral window 
function and the amplitude spectrum of the data. If a signal was found 
to be present at a significant level and supposed intrinsic, it was 
fitted to the data, and its amplitude, frequency and phase were improved 
to obtain an optimal fit in each filter. Then, a signal-to-noise 
weighted mean frequency was computed for the data in each filter, 
adopted for all, and only the amplitudes and phases were optimized.

Consequently, this variation was subtracted from the measurements (we 
call this procedure "prewhitening") and the residual amplitude spectrum 
was computed and examined. If more periodic signals were present, their 
parameters were optimized together with those of the ones previously 
detected. Once no significant variation was left in the residuals, the 
analysis was stopped. A signal was considered significant when it 
exceeded a S/N ratio of four, following the empirical recommendation by 
Breger et al.\ (1993).

\subsection{$\kappa$ Sco}

The spectral window, as well as original and prewhitened Fourier 
amplitude spectra of our $\kappa$ Sco measurements are shown in Fig.\ 2. 
As the spectral window function of our light curves is for obvious 
reasons virtually the same, we only show it once in this paper. For the 
present example, we calculated it as the Fourier Transform of a single 
noise-free sinusoid with a frequency of 5.003\,\cd and an amplitude of 
5.2 mmag. This is the same as the strongest signal in our $v$ filter 
data, that were also chosen to illustrate the prewhitening process.

\begin{figure}
\includegraphics[width=83mm]{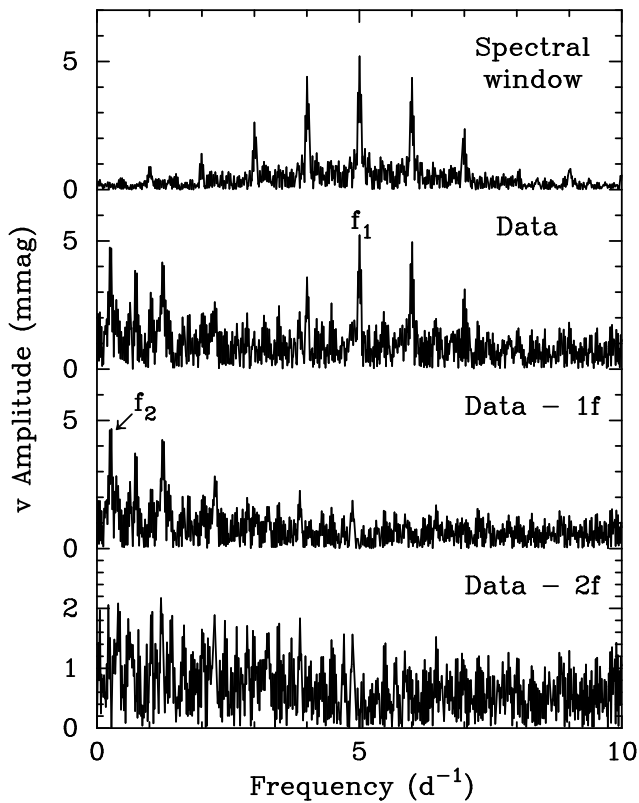}
\caption{Uppermost panel: spectral window of our $v$ filter data of
$\kappa$ Sco. Second panel: amplitude spectrum of our data. Third panel:
residual amplitude spectrum after prewhitening the dominant variation.
Lower panel: residuals after prewhitening two frequencies.}
\end{figure}

We were able to detect two statistically significant signals in our light
curves and present the frequency solution in Table 1. It should be noted
that there are some alias ambiguities concerning the second signal at low
frequency. We decided to choose the frequency value that represented the
variability in the combined $uvy$ data best; competing frequencies were at
0.244, 1.245 and 1.276\,\cd. The errors in the parameters listed in Table
1 were computed the analytical formulae by Montgomery \& O'Donoghue
(1999). These are known to underestimate the real accuracy of the
frequency solution, but give an idea of the order of magnitude of the
errors.

\begin{table}
\begin{center}
\caption{Frequencies present in the light curves of $\kappa$ Sco. The
formal errors on the frequencies are $\pm 0.001$\,\cd; on the amplitudes
they are $\pm 0.5$ mmag in v and y and $\pm 0.6$ mmag in u. The S/N ratio
is specified for the v filter data. The signals are in phase within the
three filters within the observational errors.}
\begin{tabular}{lccccc}
\hline
ID & Frequency & \multicolumn{3}{c}{Amplitude} & $S/N$ \\
 &  & u & v & y & \\
 & (\cd) & (mmag) & (mmag) & (mmag) & \\
\hline
$f_1$ & 5.003 & 7.4 & 5.4 & 5.2 & 8.0 \\
$f_2$ & 0.274 & 6.3 & 4.4 & 3.0 & 4.7 \\
\hline
\end{tabular}
\end{center}
\end{table}

$\kappa$ Sco is known to be a spectroscopic binary system with two 
B-type components in a 195-d orbit, and the primary is responsible for 
the pulsational variability (Harmanec, Uytterhoeven, \& Aerts 2004). 
Uytterhoeven et al.\ (2005) spectroscopically determined $T_{\rm 
eff}=25000\pm2000$\,K for the primary component, $T_{\rm 
eff}=20000\pm2000$\,K for the secondary, as well as log $g = 3.8\pm0.2$ 
for both. As this large temperature difference is not possible assuming 
coeval evolution, we only take these values as starting points for 
placing the two components in the HR Diagram.

Other available constraints comprise the HIPPARCOS parallax of the 
system ($6.75\pm0.17$\,mas, van Leeuwen 2007), translating into 
$M_v=-3.48\pm0.06$ and a system luminosity log $L = 4.23\pm0.03$ using 
the bolometric corrections by Flower (1996) and the uncertainties in 
$T_{\rm eff}$. The spectroscopic mass ratio is 1.1 (Harmanec, 
Uytterhoeven, \& Aerts 2004), and the primary component oscillates with 
a frequency of 5.003\cd (Table 1).

We computed model evolutionary tracks with the Warsaw-New Jersey stellar 
evolution code (e.g., see Pamyatnykh et al.\ (1998) for a description) 
to bring all these constraints into agreement. The models were 
calculated using OP opacities and the Asplund et al.\ (2004) mixture. An 
overall metal abundance $Z=0.012$ and a hydrogen abundance of $X=0.7$ 
have been adopted for all models, and no convective core overshooting 
was used. Zero-Age Main Sequence (ZAMS) rotational velocities of 115 
\kms and 195 \kms, were adopted to match the observed $v \sin i$ values 
of 100 and 170 \kms for the primary and secondary, respectively, during 
somewhat advanced main sequence evolutionary stages.

We found that model pairs in a rather narrow range of parameters 
($M_1=10.2\pm0.3 M_{\sun}$, $M_2=9.3\pm0.3 M_{\sun}$, log 
$g_1=3.83\pm0.06, \log g_2 - \log g_1 = 0.09\pm0.02$, log $T_{\rm 
eff,1}=4.365\pm0.015$, log $T_{\rm eff,2} - \log T_{\rm eff,1} 
=-0.008\pm0.005$, log (age) $=7.18\pm 0.05$) would fit all the 
observational constraints satisfactorily and within the errors. The 
orbital solution by Harmanec, Uytterhoeven \& Aerts (2004) implied 
$M_1\sin^3i=10.69 M_{\sun}$ and $M_2\sin^3i=9.64 M_{\sun}$, which 
yields $i<77\degr$.

To identify the spherical degree of the oscillation we have computed
theoretical $uvy$ amplitudes of models of $M=10.2\pm0.3 M_{\sun}$ and
log $T_{\rm eff,1}=4.365\pm0.015$ following the approach by Balona \&
Evers (1999). We calculated theoretical uvy amplitudes for the modes whose
frequencies were in the observed domain along this model sequence. The
resulting theoretical amplitudes in the three passbands were normalized to
that in the u filter, and the average amplitude ratios and their rms
errors were calculated. Given the small temperature difference between the
primary and secondary component, the falsification of the amplitude ratios
due to the secondary flux is insignificant compared to the observational
errors and hence was not taken into account. A comparison between the
observed and theoretical amplitude ratios is shown in Fig.\ 3.

\begin{figure}
\includegraphics[width=89mm]{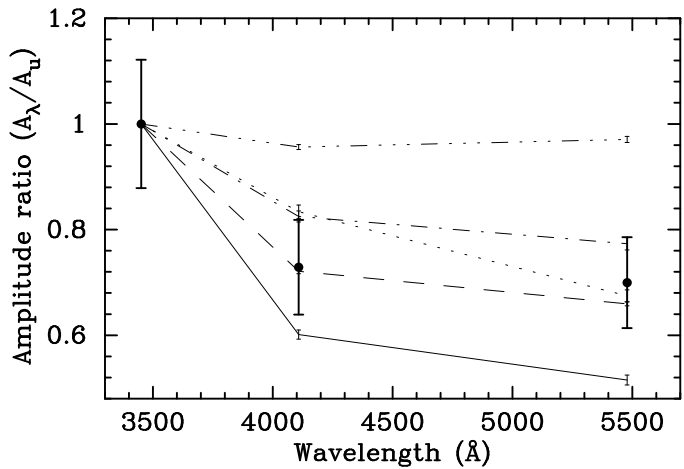}
\caption{A comparison of the observed amplitudes of $\kappa$~Sco in the
different filters with theoretical predictions of pulsational $uvy$
amplitude ratios, normalized to unity at $u$. The filled circles with
error bars are the observed amplitude ratios. The full lines are
theoretical predictions for radial modes, the dashed lines for dipole
modes, the dashed-dotted lines for quadrupole modes, the dotted lines for
$\ell=3$ modes, and the dashed-triple-dotted lines are for $\ell=4$. The
thin error bars denote the uncertainties in the theoretical amplitude
ratios.}
\end{figure}

It is implied that the $\beta$~Cephei oscillation is most consistent 
with a dipole mode. The dominant pulsation mode of the primary component 
has been spectroscopically identified as either $(l, m) = (2, -1)$ or 
$(1, -1)$ (Uytterhoeven et al.\ 2005). Our results point towards the 
latter possibility, which is also more favourable when considering the 
geometrical aspect of the mode (e.g., see Pesnell 1985) assuming the 
primary's rotation axis is normal to the orbital plane.

We now turn to the 0.274\,\cd variation. A signal in this frequency 
domain has been observed before, both in space photometry (Cuypers, 
Buzasi \& Uytterhoeven 2004) and in spectroscopy (Uytterhoeven et al.\ 
2005). The spectroscopic observations were interpreted as a rotational 
modulation. However, its photometric amplitudes with respect to 
wavelength (Table 1) are also consistent with pulsation. On the other 
hand, the frequency reported by Uytterhoeven et al.\ (2005) is somewhat 
lower than the one detected in our and the space photometry. In fact, it 
corresponds to one of the aliases (0.244\,\cd) mentioned before. With a 
radius of $R=6.45\pm0.55 R_{\sun}$ resulting from the model 
computations described above and $v_{\rm rot}= 110$\,\kms, we obtain a 
rotation period of $2.97\pm0.25$\,d for the primary component. This is 
somewhat shorter than the observed photometric and spectroscopic 
periods, but given the uncertainties in the preceding procedures, we 
cannot unambiguously distinguish between a rotational modulation or a 
g-mode oscillation with a period very close to the rotational one.

It is interesting to examine the photometric data of $\kappa$~Sco for 
possible eclipses. Indeed, two nights (HJD 2441120 and 2441122) by Lomb 
\& Shobbrook (1975) show magnitude zeropoints 0.024 mag fainter than the 
rest. However, data from the same authors from different seasons, but at 
the same orbital phase according to the spectroscopic 195.65-d orbital 
period (Uytterhoeven et al.\ 2001), do not show fainter mean magnitudes 
and therefore rather indicate an observational problem in those two 
nights. In addition, in one of these two nights $\lambda$ Sco was also 
observed by these authors. This night shows a similar mean magnitude 
change in the $\lambda$ Sco data. Consequently, there is no evidence for 
possible eclipses of $\kappa$~Sco in the photometry available to us.

\subsection{$\lambda$ Sco}

As can already be discerned in Fig.\ 1, in the night of HJD 
2455760 (17/18 July 2011) $\lambda$ Sco underwent part of a shallow 
eclipse. To determine the frequencies of its intrinsic light variations, 
we therefore discarded the data affected by the eclipse for the time 
being. Furthermore, after making sure that any variability had the same 
amplitude within the errors in both filters, we combined those data 
after setting their mean to zero. The frequency analysis of the 
remaining combined $vy$ data is presented in Fig.\ 4, and the results, 
with the amplitudes individually determined for the different filters, 
in Table 2. The signals are in phase within the two filters within the 
observational errors. Given the lack of $u$ data for $\lambda$~Sco, we 
have no means to attempt a photometric identification of the dominant 
oscillation mode, despite its rather large amplitude.

\begin{figure}
\includegraphics[width=83mm]{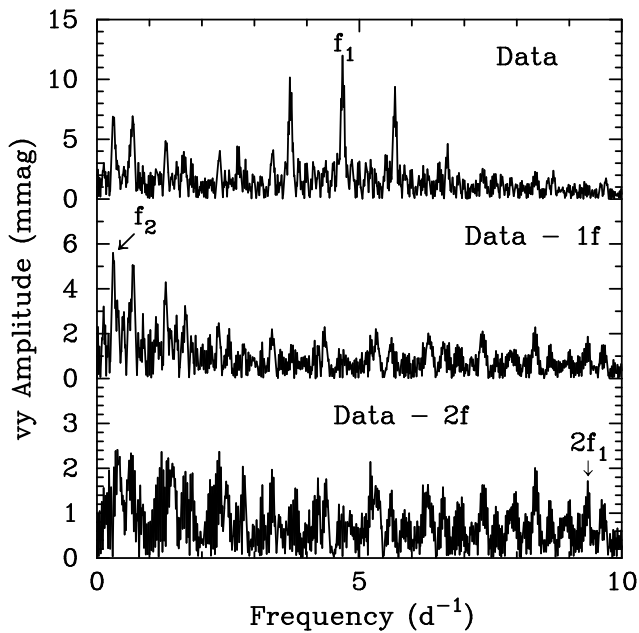}
\caption{Upper panel: amplitude spectrum of our data of $\lambda$ Sco with
the eclipse discarded. Second panel: residual amplitude spectrum after
prewhitening the dominant variation. Lower panel: residuals after
prewhitening two frequencies. A peak at the harmonic of the first
variation is present.}
\end{figure}

The longer-period signal we found in the light variations has never been 
reported before. It appears convincingly present in our data, but we 
lack observational constraints to interpret it. However, one statement 
can be made: its frequency is close to, but significantly different 
from, the first harmonic of the orbital frequency of the close binary, 
which argues against a tidal origin.

$\lambda$~Sco is a known triple system, consisting of a close double 
with an orbital period of 5.9525\,d and of a tertiary star in a $\approx 
3$-yr orbit. The primary is the $\beta$~Cephei star, the secondary has a 
mass of about 1/6 that of the primary, whereas the tertiary again is a 
B-type star with a mass of about 76\% of the primary. We refer to 
Uytterhoeven et al.\ (2004, hereinafter U04) and Tango et al.\ (2006) 
for in-depth 
studies of the triple system. The masses and effective temperatures 
listed in Table 5 of Tango et al.\ (2006) lead to 
$R=5.2\pm1.7 R_{\sun}$ for the primary and $R=5.3\pm1.3 R_{\sun}$ 
for the tertiary, respectively. Hence, the pulsation constant for the 
$\beta$~Cephei oscillation of the primary would be $Q=0.06\pm0.02$, 
implying a low-order g mode, which is rather unusual for $\beta$~Cephei 
stars (cf.\ Stankov \& Handler 2005). Additionally, Tango et al.\ (2006) 
reported a smaller distance of the $\lambda$~Sco system obtained from 
their combined interferometric and spectroscopic analysis compared to 
that from the HIPPARCOS parallax. This raises the suspicion that the 
component masses, and hence the luminosities, of the components of the 
$\lambda$~Sco system have previously been underestimated.

Consequently, we now apply the same methodology as for $\kappa$~Sco, 
namely trying to find stellar model pairs of the same age in the 
effective temperature range implied by spectroscopy that are consistent 
with the system's parallax. In this procedure, we neglect the luminosity 
of the secondary as it is insignificant within the errors. The 
luminosity ratio of the tertiary to the primary is $0.38\pm0.08$ (Tango 
et al.\ 2006). With the HIPPARCOS parallax of $\lambda$ Sco 
($5.71\pm0.75$\,mas, van Leeuwen 2007) and the bolometric corrections by 
Flower (1996), it follows that the primary component must have $\log 
L_1=4.56\pm0.13$. At an effective temperature of $T_{\rm 
eff,1}=25000\pm1000$\,K (U04) this corresponds to 
a mass of $14.5\pm1.1 M_{\sun}$ according to our models, near the 
upper mass limit for the primary star inferred by Tango et al.\ (2006). 
We note that models with $M>14.5 M_{\sun}$ without overshooting are 
at the terminal-age main sequence at $T_{\rm eff}=25000$\,K. The 
pulsation constant for the $\beta$~Cephei mode of the primary component 
of $\lambda$ Sco for models between $13.4-14.5 M_{\sun}$ lie in the 
range 0.026\,d$<Q<$\,0.031\,d, in good agreement with common values for 
$\beta$~Cephei stars (Stankov \& Handler 2005). With the tertiary to 
primary mass ratio of $0.76\pm0.04$ (Tango et al.\ 2006), the tertiary 
star must have $9.6-11.6 M_{\sun}$. In this scenario, assuming coeval 
evolution and $T_{\rm eff,1}=25000\pm1000$\,K, it would however need to 
be hotter than previously estimated ($T_{\rm eff,3}=24500\pm1000$\,K).

The rotation periods of these models, assuming that the primary's 
rotation axis is normal to the orbital plane, lie in the range of 
$3.4\pm0.5$\,d, given $v\sin i\approx 150$\,\kms (see U04). This is in 
very good agreement with the low frequency detected in our photometry, 
which may therefore correspond to the rotation frequency or to a 
pulsational variation very close to it.

\begin{table}
\begin{center}
\caption{Frequencies present in the light curves of $\lambda$ Sco. The
formal errors on the amplitudes are $\pm 0.6$ mmag. The S/N ratio is
specified for the combined vy filter data.}
\begin{tabular}{lcccc}
\hline
ID & Frequency & \multicolumn{2}{c}{Amplitude} & $S/N$ \\
 &  & v & y & \\
 & (\cd) & (mmag) & (mmag) & \\
\hline
$f_1$ & $4.6784 \pm 0.0006$ & 12.0 & 12.4 & 15.7 \\
$f_2$ & $0.299 \pm 0.001$ & 5.3 & 6.2 & 5.2 \\
\hline
\end{tabular}
\end{center}
\end{table}

With the light curve fit from Table 2, we predicted the light curve for 
the night of the eclipse, subtracted it from the data, and show the 
rectified eclipse light curve in Fig.\ 5. The eclipse is a transit and 
about 0.03 mag deep. Eclipses of $\lambda$~Sco have been seen before 
(Shobbrook \& Lomb 1972, hereinafter SL72, Bruntt \& Buzasi 2006) and 
are due to the closer pair in the $\lambda$~Sco triple system (with an 
orbital period of 5.9525\,d according to U04). Folding our data with 
this orbital period provides no evidence for the detection of a 
secondary eclipse; our data also do not cover the predicted times of 
further eclipses. From the primary's and tertiary's parameters inferred 
above, we thus obtain $R_1=8.8\pm1.2 R_{\sun}$, $R_2=1.5\pm0.2 
R_{\sun}$, and $R_3=4.7\pm1.0 R_{\sun}$. The secondary would then be a 
$M_2=2.0\pm0.2 M_{\sun}$ star at the beginning of main sequence 
evolution.

\begin{figure}
\includegraphics[width=83mm]{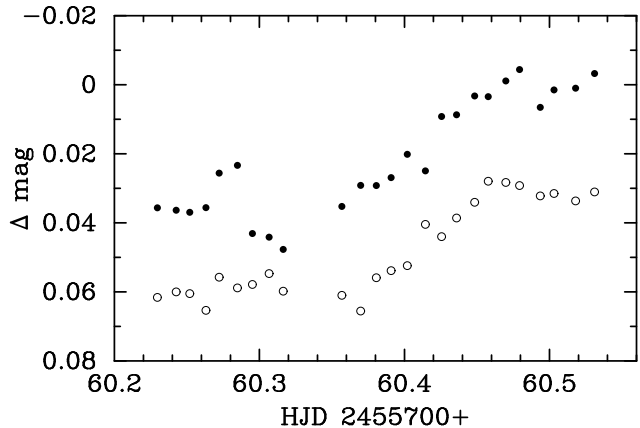}
\caption{Eclipse of $\lambda$~Sco observed on the night of 17/18 July
2011. The filled circles are the $v$ filter data, the open circles the
measurements in $y$.}
\end{figure}

The light curves by Bruntt \& Buzasi (2006) allow to estimate that the 
eclipses of the primary of $\lambda$~Sco by the secondary last about 
8.5\,hr from mid-ingress to mid-egress. Unfortunately, the figures in 
this paper do not permit unambiguous eclipse timing determinations. We 
are therefore left with the partial eclipses in SL72's and our data, as 
well as with two eclipses seen in the HIPPARCOS photometry of the star. 
The times of central eclipse determined from these data are listed in 
Table 3.

\begin{table}
\begin{center}
\caption{Times of central eclipse of the primary of $\lambda$~Sco and 
comparison with the spectroscopic ephemeris of the close pair 
by U04.}
\begin{tabular}{cccl}
\hline
Time & E & $O-C$ & Reference\\
(HJD $-$ 2400000) & & (d) & \\
\hline
$40690.05\pm0.02$ & $-1675$ & $0.98\pm0.02$ & SL72\\
$48165.68\pm0.04$ & $-419$ & $0.27\pm0.04$ & ESA (1997)\\
$48659.74\pm0.03$ & $-336$ & $0.27\pm0.03$ & ESA (1997)\\
$55760.24\pm0.02$ & 857 & $-0.56\pm0.02$ & this paper\\
\hline
\end{tabular}
\end{center}
\end{table}

Using the spectroscopic ephemeris of the close binary in $\lambda$~Sco 
(U04), we can estimate an epoch zero of a potential eclipse of HJD 
$2450659.51\pm0.06$, and determine $(O-C)$ values of the times of 
central eclipse for the 5.9525-d orbital period. These deviate 
considerably from the ephemeris (cf.\ Table 3). The light time effect 
caused by the tertiary component cannot be the cause of these $(O-C)$ 
deviations because its maximum range can only be 0.03\,d.

Consequently, we re-analysed the radial velocities by U04 with {\tt 
Period04}. A fit with the two orbital periods, the first harmonic of the 
shorter orbital variation and with the dominant pulsation frequency 
resulted in an orbital period of the close system of 
$5.95242\pm0.00004$\,d, within $3\sigma$ of U04. The formal error quoted 
here (cf.\ Montgomery \& O'Donoghue 1999) is the same as given in U04. 
However, the data set is unevenly distributed: 74\% of the radial 
velocities originate from only nine nights out of a total of 77, and the 
first two nights are separated by seven years from the remainder that 
also span seven years.

We therefore summed the data into bins separated by one pulsation period 
and repeated the frequency analysis (omitting the pulsation frequency). 
This resulted in an orbital period of the closer pair of 
$5.9522\pm0.0002$\,d, and twice this uncertainty when not considering 
the first two nights. We therefore argue that the discrepancy between 
the spectroscopic orbital period of U04 and our eclipse timings is 
simply due to an earlier underestimate of the observational errors. 
Consequently we propose a refined value for the orbital period of the 
closer pair of $5.95189\pm0.00003$\,d, to bring the photometric and 
spectroscopic measurements into agreement.

\subsection{$\mu^1$ Sco}

Given its brightness, its rather large light amplitude, and discovery as a
spectroscopic binary already in the 19th century, there is rather little
literature data for this eclipsing binary. On the other hand, a recent
comprehensive study of this system was published by van Antwerpen \& Moon
(2010), wherefore it is of little point to repeat these authors' work.  
However, we are interested in obtaining good binary light curve fits
because the primary component is in the $\beta$ Cephei instability strip
and the secondary is in the domain of SPB stars. 

\begin{figure}
\includegraphics[width=83mm]{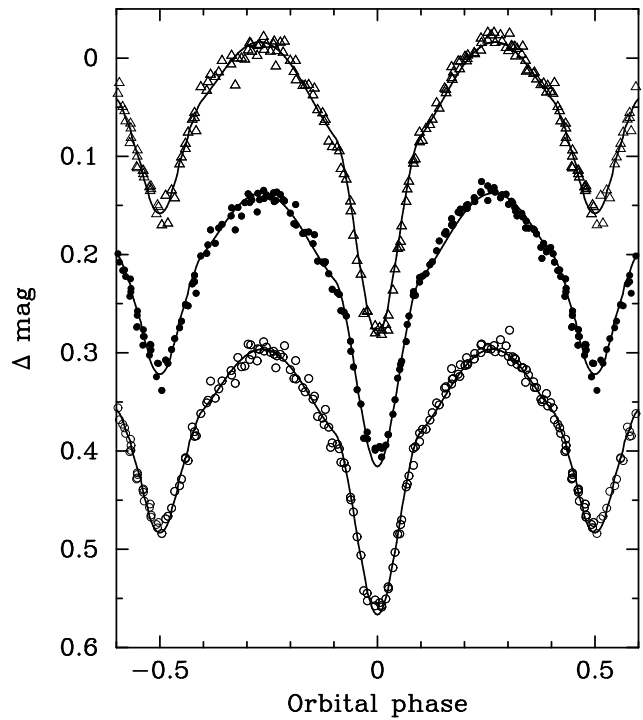}
\caption{Phased light curves of $\mu^1$ Sco according to the
ephemeris by van Antwerpen \& Moon (2010). The triangles are our $u$
measurements, the filled circles are our $v$ filter data, the open circles
our measurements in $y$. The lines are fits to the data.}
\end{figure}

We therefore applied the PHysics Of Eclipsing BinariEs (PHOEBE) code 
(Pr\v{s}a \& Zwitter 2005) to our light curves, using the binary 
parameters from van Antwerpen \& Moon (2010) as starting values. It is 
implied that the ephemeris given by these authors is still valid for our 
data (Fig.\ 6). However, according to our knowledge our measurements 
yielded the first photometric data of $\mu^1$ Sco in the ultraviolet. 
Indeed, irradiation of the cooler secondary (implemented in PHOEBE) had 
to be included to provide a fit our u data within the observational and 
analysis errors - as already predicted by Ruci{\'n}ski (1970).

Subtracting the light curve solution leaves non-white residuals with the 
two most interesting modulation frequencies around 0.123 and 8.07 c/d 
and about 3 mmag amplitude that are visible in all three filters' data. 
These might correspond to intrinsic variability of at least one of the 
binary components, but do not reach the required $S/N\geq4$ to be 
considered as detected. Future BRITE data will enable us to investigate 
this possibility deeper.

\subsection{$\zeta^1$ Sco}

The light variations of this star are only discernible from night to
night. Consequently, we show nightly averaged magnitude and colour
variations of $\zeta^1$ Sco in Fig.\ 7. A period search in these data is
(as to be expected from the light curves) not very revealing and implies
a time scale of the variations of about two or four weeks. The latter is
compatible with two of the variability periods of $\zeta^1$ Sco proposed
by Sterken, de Groot \& van Genderen (1997), but our data do not fold
well with either of these two periods. There is also little colour
variation associated with the light changes, in particular in $(u-y)$,
which can be taken as an argument against normal-mode pulsations.

\begin{figure}
\includegraphics[width=83mm]{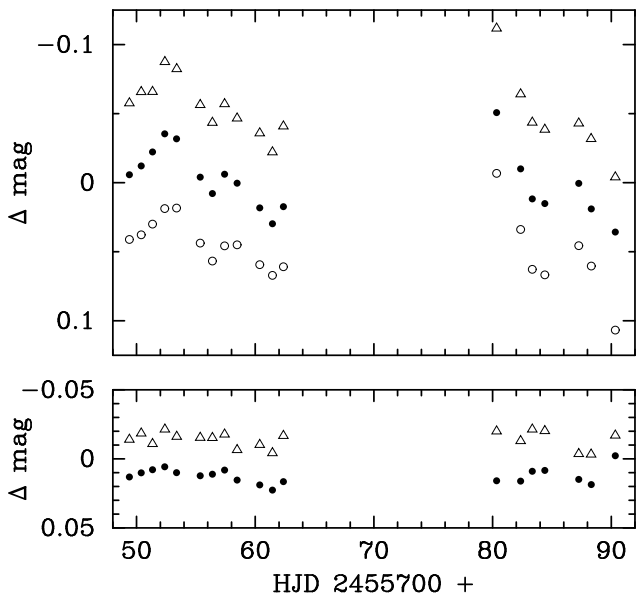}
\caption{Upper panel: our light curves of $\zeta^1$ Sco. The triangles are
our $u$ measurements, the filled circles are our $v$ filter data, the open
circles our measurements in $y$. Lower panel: the $u-y$ (triangles) and
$v-y$ (filled circles) colour variations.}
\end{figure}

$\zeta^1$ Sco is a very massive ($M \geq 36 M_{\sun}$) supergiant
(Clark et al.\ 2012) and shows evidence for wind clumping (Prinja \& Massa
2010). As a consequence of its high mass and luminosity, strange modes may
excited (Gautschy 2009). To us, these two possibilities appear to be more
promising to explain the observed variability of the star, as opposed to
normal mode pulsations.

\section{Summary and conclusions}

In anticipation of the BRITE-Constellation space mission, we have carried
out time-resolved multicolour photometry of six diverse bright early-type
stars in Scorpius, a sky region certainly interesting to be studied with
BRITE. The magnitude range and temporal behaviour of those stars is also
what is to be expected to be close to what a typical BRITE field will
contain.

We have detected one long and one short period for each of the two known 
$\beta$ Cephei stars $\kappa$ and $\lambda$ Sco. We observed part of an 
eclipse of the latter star and found a longer period in its light 
variation that has never been reported before. Due to the lack of 
ultraviolet data, we cannot provide a mode identification for $\lambda$ 
Sco, but for $\kappa$ Sco we were able to constrain $l \leq 1$ for the 
dominant $\beta$ Cephei pulsation mode. Together with spectroscopic 
results from the literature, this makes this mode most likely $(l, m) = 
(1, -1)$. The $uvy$ amplitudes of the longer period of $\kappa$ Sco, are 
consistent with pulsation. However, a rotational origin can also not be 
ruled out, and a hypothesized longer-period pulsation would have a 
frequency very close to the rotational one.

A rotational modulation or g-mode pulsation may also be present in 
$\lambda$ Sco on top of the known $\beta$ Cephei oscillation. Previous 
analyses of this triple system are not in good agreement with its 
HIPPARCOS parallax and the pulsational behaviour of the primary. 
Therefore we argued that the components of this system are some 30\% 
more massive than previously thought. We also noticed that the 
spectroscopic ephemeris of the close binary in the system does not 
predict observed eclipse timings correctly, and proposed a revised value 
for the orbital period.

We obtained full orbital coverage of the light curve of the known 
eclipsing binary $\mu^1$ Sco, and found its variability consistent with 
the latest ephemeris in the literature. We presented the first 
ultraviolet light curves of the system, which could only be 
satisfactorily modelled by including (long ago predicted) irradiation 
effects. There are hints of additional variability that may be 
pulsational in origin, which however remains to be revealed with 
BRITE-Constellation. Finally, our observations of the blue 
super/hypergiant $\zeta^1$ Sco showed variability on time scales of 
weeks, that we prefer to interpret in terms of wind variations or 
strange mode oscillations rather than normal mode pulsations.

Given the modest size of the present study, the number of new results on 
these bright stars is astounding. Part of the reason may be that ground 
based photometric studies of the brightest stars in the sky are 
technically difficult, wherefore comparatively less about them is known 
than for fainter stars. Consequently, we are eagerly looking forward to 
the new scientific results to be delivered by BRITE-Constellation.

\begin{acknowledgements}

This work was supported by the Polish NCN grant 2011/01/B/ST9/05448 and 
the BRITE PMN grant 2011/01/M/ST9/05914. We thank Bob Shobbrook for 
digitizing the archival observations of $\kappa$ and $\lambda$ Sco, 
Stefan Mochnacki for helpful discussions about binary star light curve 
modelling, and Katrien Uytterhoeven for providing her spectroscopic data 
of $\lambda$ Sco.

\end{acknowledgements}

\end{document}